# Dumbbell Stanane: A large-gap quantum spin Hall insulator

Xin Chen, Linyang Li, Mingwen Zhao*

*School of Physics and State Key Laboratory of Crystal Materials, Shandong University, Jinan, Shandong, 250100, China*

**Abstract:** Quantum spin Hall (QSH) effect is quite promising for applications in spintronics and quantum computations, but presently can only be achieved at ultralow temperature. Searching for large-gap QSH insulators is the key to increase the operating temperature. Using first-principles calculations, we demonstrate that the stable hydrogenated stanene with a dumbbell-like structure (DB stanane) has large topological nontrivial band gaps of 312 meV ($\Gamma$ point) and 160 meV for bulk characterized by a topological invariant of $Z_2=1$, due to the s-$p_{xy}$ band inversion. Helical gapless edge states appear in the nanoribbon structures with high Fermi velocity comparable to that of graphene. The nontrivial topological states are robust against the substrate effects. The realization of this material is a feasible solution for applications of QSH effect at room temperature and beneficial to the fabrication of high-speed spintronics devices.

Corresponding author, E-mail: zmw@sdu.edu.cn

## Introductions

Topological insulators (TIs) have drawn increasing interests of fundamental condense-matter physics and material science owing to an insulating bulk and gapless surface or edge states protected by time-reversal symmetry[1-4]. The low-energy scattering of the edge states is enjoined by the time-reversal symmetry, leading to the dissipationless transport edge channels, which is promising for applications in spintronics and quantum computations[5-9]. Two-dimensional (2D) TIs, also known as quantum spin Hall (QSH) insulators, was first proposed in graphene[3]. However, subsequent works indicated that the nontrivial band gap opened due to spin-orbit coupling (SOC) is unobservably small (~$10^{-3}$ meV) in graphene, which limits the operating regime to unrealistically low temperatures. Although many materials have been theoretically predicted to be QSH insulators, only the HgTe/CdTe[10] and InAs/GaSb[11] quantum wells are verified by transport experiments. Extremely low temperature condition is still required due to their small bulk gap. Searching for large-gap QSH insulators is the key to increase the operating temperature. Remarkable works have been done for large-gap QSH insulators, such as metal-organic frameworks[12-15] and those containing heavy metal atoms[16-20] or under a tensile strain[21].

The 2D materials of group-IV elements have attracted attentions for their rich electronic properties[22-24]. Low-buckled (LB) silicene, germanene and stanene, as well as their halogenides and hydrogenides[25, 26] have been proposed as alternates of graphene to achieve QSH effect at high temperature. Compared with graphene, the SOC strength in these materials is greatly enhanced and the bulk gaps are larger than the thermal motion energy at room temperature (~26 meV). However, the relatively weak π-π bonding of the LB configurations may fail to stabilize the buckled configuration. A $2 \times 2$ restructured dumbbell (DB) structure containing fourfold-coordinated atoms, as shown in Fig. 1(a), was proposed as a stable phase of these 2D materials[27-30]. Interestingly, the recent work of Tang et al. [27] indicated that the DB stanene is a 2D QSH insulator with an inverted band gap. Although the bulk band gap in free-standing DB stanene is smaller than that of the LB stanene[25], it can be improved by the interaction with the substrate, such as reconstructed (2×2) InSb(111) surface. It is noteworthy that this DB stanene contains 40% three-fold coordinated Sn atoms, which may be chemically reactive and thus disadvantageous for the realization and practical applications.

In this contribution, using first-principles calculations, we showed that hydrogen atoms can chemically bond to the under-coordinated Sn atoms of the DB stanene, leading to a hydrogenated DB stanene ($Sn_{10}H_4$) with high energetic and dynamic stability. We named the hydrogenated stanene as DB stanane. We demonstrated that the DB stanane is a QSH insulator characterized by a topological invariant of $Z_2=1$. The bulk band gap at the Γ point is about 312 meV, much larger than that in the DB stanene grown on InSb substrate. The topological nontriviality is confirmed by the appearance of helical gapless edge states in nanoribbons. The Fermi velocity of the edge states is comparable to that of graphene. The realization of the DB stanane is a feasible solution for the applications of QSH effect at room temperature and beneficial to the fabrication of high-speed spintronics devices.

## Methods

Our first-principles calculations were performed using the plane wave basis Vienna ab initio simulation package known as VASP code[31-33], implementing the density functional theory (DFT). The electron exchange-correlation functional was treated using generalized gradient approximation (GGA) in the form proposed by Perdew, Burke, and Ernzerhof (PBE)[34]. The atomic positions were relaxed until the maximum force on each atom was less than 0.01 eV/Å. The energy cutoff of the plane waves was set to 600 eV with the energy precision of $10^{-5}$ eV. For the 2D structures, the Brillouin zone (BZ) was sampled by using an 11×11×1 Gamma-centered Monkhorst-Pack grid, while a 1×11×1 grid was used for the nanoribbon. The vacuum space was set to at least 20 Å in the calculations to minimize artificial interactions between neighboring slabs. SOC was included by a second variational procedure on a fully self-consistent basis. The phonon spectra were calculated using a supercell approach within the PHONON code[35].

## Results and Discussion

In DB stanene, as shown in Fig. 1(a), there are two types of Sn atoms: one is four-fold coordinated (denoted as Sn(α)), similar to the case of bulk crystal, another is only three-fold coordinated with a dangling bond (denoted as Sn(β)) in analogous to the case of LB stanene. All the Sn(α) atoms are on the same plane sandwiched by up- and down-planes of Sn(β). The buckling height of the DB stanene measured from the distance between the two planes of Sn(β) is about 3.41 Å. The lattice constant of the 2D hexagonal structure is 9.05 Å. These results are in agreement with those reported in previous work[27]. The stability of the DB stanene over the LB stanene composing purely of three-fold coordinated Sn atoms arises from the four-fold coordinated Sn atoms[28-30], because four-fold coordinated Sn is energetically more preferable.

Considering that Sn can form stannane ($SnH_4$) with hydrogen, we therefore try to passivate the dangling bond of Sn(β) by hydrogen atoms. The equilibrium configuration of the DB stanane ($Sn_{10}H_4$) is shown in Fig. 1(b). All the Sn atoms in this configuration are fully-coordinated. The H-Sn(β) distance is 1.74 Å, close to the value 1.72 Å in stannane molecule. The results clearly indicate that hydrogen atom chemically binds to the underlying Sn atom, in consistent with the electron localization function (ELF)[36, 37] profile shown in Fig. 1(c). The length of the Sn(α)-Sn(β) bond is about 2.87 Å, slightly shorter than that of the Sn(α)-Sn(α) bond, 2.89 Å, and close to the bond lengths in LB stanene (2.83 Å) and Tin (2.88 Å). These covalent bonds are preserved in the DB stanane as indicate by ELF profile. The distance between the up and down Sn planes is 3.52 Å, slightly longer than that of the DB stanene, 3.41 Å. The lattice constant is compressed to 8.98 Å, compared to DB stanene.

We calculated the formation energy of DB stanane by the difference between the total energy of DB stanane and the sum of the total energies of DB stanene and isolated hydrogen atoms. Such atomic hydrogen can be achieved in hydrogen plasma environment as used in hydrogenating graphene[38, 39]. The formation energy is about -0.55 eV/atom. The negative formation energy implies the energetic superiority to the DB stanene. The energetic stability of the DB stanane can also be hinted by the following hypothetical reaction: $10SnH_4 \rightarrow Sn_{10}H_4 + 9H_2$. Our calculations showed that this reaction is exothermal with energy release of 0.20 eV/atom. The dynamic stability is confirmed by the phonon spectrum calculated along the highly symmetric directions in the BZ, as shown in Fig. 1(c). There are no modes with imaginary frequencies in the spectrum and the film is therefore expected to be dynamically stable.

The energetic favorability of the DB configuration over the LB configuration has also been demonstrated in other 2D materials[27,29,40-43], such as silicene and germanene. Incidentally, the DB silicene and DB germanene have been proved to be the most stable structures in corresponding free-standing DB-based honeycomb derivatives involved in the works of Cahangirov et al.[29] and Özcelik et al[43]. There is already some experimental evidence of the plausibility of these DB configurations. For example, the $\sqrt{3} \times \sqrt{3}$ silicene multi-layers grown on an Ag(111) substrate[44,45,46] can be reproduced by stacking DB-based silicene derivatives[29,47]. The 2×2 superstructure of germanene grown an Al(111) substrate resemble the DB germanene very well[48]. Hydrogenation of silicene has also been achieved in a recent Qiu et al.[49], These experimental progresses imply the plausibility of DB stanene, although the synthesis of this new 2D material remains challenging at present.

The electronic structures of the DB stanane were then calculated without considering SOC. The electron band structures shown in Fig. 2 (a) clearly indicate that it is a gapless semiconductor with the valence and conduction bands meeting at a single point. The electronic states in the vicinity of the Fermi level originate from different atomic orbitals. The valence band maximum (VBM) and the conduction band minimum (CBM) which are energetically degenerated at the $\Gamma$ point arise mainly from the $p_{x,y}$ atomic orbitals of the Sn($\alpha$) atoms, as indicated by the orbital-resolved electron density of states (DOS) and band structures. The density profiles of the electron wavefunctions of the VBM and CBM exhibit the features of the binding states of $p_{x,y}$ atomic orbitals. The degeneracy of the two $p_{x,y}$ bands at the $\Gamma$ point is a direct consequence of the $C_{3v}$ symmetry of the lattice. The valence band nearest to the VBM (denoted as VBM-1), however, comes from the s-orbital of the Sn($\alpha$) and $p_z$-orbital of Sn($\beta$) (see the Supplementary Information). Such band alignment has also been reported for the halogenated germanene[25, 26]. The gapless feature differs significantly from the case of DB stanene which is an indirect-band-gap semiconductor. For the DB stanene, the direct band gap at the $\Gamma$ point is about 133 meV, accompanied by a global indirect band gap of 14 meV from our calculations. We then turned on the SOC in the DFT calculations. A band gap of 312 meV is opened up at the $\Gamma$ point of the DB stanane, due to SOC effect. The global indirect band gap is about 160 meV, as shown in Fig. 2(b). These SOC gaps are much larger than those in free-standing DB stanene, which are 43 meV ($\Gamma$ point) and 22 meV (indirect), and the values of DB stanene grown on a reconstructed (2×2) InSb(111) surface, which are 161 meV ($\Gamma$ point) and 40 meV (global) [27]. The large SOC gap in the DB stanane is quite promising for achieving QSH effect at high temperature.

To determine the topological features, we start from the alignment of the energy levels of DB stanene and DB stanane. We label the electronic states arising from s, $p_{x,y}$, $p_z$ atomic orbitals of Sn as $|s^{\pm}>$, $|p_{x,y}^{\pm}>$ and $|p_z^{\pm}>$, respectively. The superscript $(+,-)$ denotes the parity of the states. For the pristine DB stanene, the electronic states near the Fermi level turn out to be $|p_{x,y}^{+}>$ and $|s^{-}>$, as shown in Fig. 3(a). The $|s^{-}>$ state is above the $|p_{x,y}^{+}>$ state in energy in DB stanene, but the order is reversed in the DB stanane, as shown Fig. 3(b). As the SOC effect was taken into account, the degeneracy of the partial-occupied $|p_{x,y}^{+}>$ state is lifted in the DB stanane, leading to a nontrivial band gap at the Fermi level. It is noteworthy that the $p_z$ state which contributes to the band inversion of the DB stanene moves away from the Fermi level due to hydrogenation. This implies that the origins of the topological nontrivial state in the DB stanene and DB stanane are different.

The topological nontriviality of the DB stanane can be confirmed using two strategies. One is the nonzero topological invariant $Z_2$, another is the gapless helical edge states in the nanoribbons. We calculated the $Z_2$ topological invariant using the parity criteria proposed by Fu and Kane[50]. For the lattice with inversion symmetry, the $Z_2$ index can be deduced from the knowledge of the parities of the four time-reversal and parity invariant points at BZ, without having to know about the global properties of the energy bands. For the honeycomb lattice of the DB stanane, the four time-reversal invariant momenta occur at $\Gamma$ and three M points which are $\Gamma(0,0)$, $M_1(0,1/2)$, $M_2(1/2,0)$, $M_3(1/2,1/2)$, respectively. The $Z_2$ invariant $\nu$ is defined by

$$(-1)^{\nu} = \prod_{i} \delta_i \text{ with } \delta_i = \prod_{m=1}^{N} \xi_{2m}(\Gamma_i)$$

for 2N occupied bands. $\xi_{2m}(\Gamma_i) = \pm 1$ is the parity eigenvalue of the *2m*-th occupied energy band at the time-reversal invariant momentum $\Gamma_i$. Our first-principles calculations showed $\delta_i$ has the values of (−), (+), (+), and (+) at the four time-reversal momenta. The topological invariant is therefore $Z_2=1$, indicating that DB stanane is a topological insulator. Such s-p-type band inversion mechanism[52] has also been reported in HgTe quantum well[7] and fluorinated stanene[25].

To reveal the existence of helical gapless edge states in the DB stanane, we considered an armchair-edged nanoribbon, without loss of universality, as is shown in Fig. 4(a). The edge Sn atoms are hydrogenated to eliminate the dangling bonds[26, 52]. The width of the nanoribbon, 10.6 nm, is large enough to avoid interactions between the edge states of the two sides. The band structure of the nanoribbon is shown in Fig. 4 (b). We can see explicitly that the gapless edge states (red lines) appear within the bulk gap and cross linearly at the $\Gamma$ point, which proves the topological nontribal property of the bulk gap. The Fermi velocity of the edge states is about $7.4\times10^5$m/s, comparable to the value $8.46\times10^5$ m/s in graphene[41, 42], both of which are larger than that of $5.0 \times 10^5$ m/s in GaBiCl$_2$ quantum well[13] and $5.5\times10^5$ m/s in HgTe/CdTe quantum well.[1] This value is also remarkable larger than that of $3\times10^4$ m/s in InAs/GaSb quantum well[11]. High Fermi velocity in the edge states of the DB stanane is beneficial to the fabrication of high-speed spintronics devices.

Substrates are evitable in the fabrication of devices. The coupling between the substrate and the TIs deposited on them may destroy the topological nontriviality or increase the SOC band gap[20, 27]. For example, the interaction between silicene and semiconducting substrate induces a trivial band gap at the Dirac point of silicene, making it a trivial semiconductor[53-55]. On the other hand, when DB stanene is deposited on a reconstructed (2×2) InSb(111) surface, the nontrivial band gap at the Γ point increases to 161 meV[27]. The recent work of Zhou et al. showed that the hexagonal lattice of Bi atoms grown on the Si(111) surface functionalized with one-third monolayer halogen atoms exhibit isolated QSH state with an energy gap as larger as ~0.8 eV, due to a substrate-orbital-filtering effect[20]. This opened a new and exciting avenue for exploration of large-gap topological surface/interface states. To reveal the effect of substrate on the topological nontriviality of the DB stanane, we considered a superstructure of DB stanane on a ($\sqrt{13} \times \sqrt{13}$) h-BN substrate, as shown in Fig. 5(a). The lattice mismatch between DB stanane and h-BN substrate is less than 1%. The optimized interlayer space between stanane and substrate is about 2.3 Å, and the binding energy between them is only 0.065 eV/atom. First-principles calculations showed that electronic band lines of the free-standing DB stanane in the region near the Fermi level are well preserved in the superstructure, as shown in Fig. 5(b) and (c). The SOC gaps are slightly affected by the h-BN substrate. This implies that h-BN may be a suitable substrate for the DB stanane in practical application.

## Conclusions

In summary, using first-principles calculations, we demonstrate that the hydrogenating DB stanene can improve not only the stability but also the SOC band gap of the QSH insulator. The bulk band gaps of the DB stanane can be as large as 312 meV (Γ point) and 160 meV (global), both of which are much larger than the values of free-standing DB stanene and the DB stanene grown on InSb substrate. The DB stanane is identified as a 2D topological insulator with a topological invariant of $Z_2 =1$. The Fermi velocity of the helical edge states $7.4\times10^5$ m/s is comparable to that in graphene. Those properties are beneficial for achieving QSH effect at high temperature as well as the fabrication of high-speed spintronics devices.

## Acknowledgements

This work is supported by the National Basic Research Program of China (No. 2012CB932302), the National Natural Science Foundation of China (No. 91221101), the 111 project (no. B13209), and the National Super Computing Centre in Jinan.

# References

1 X. L. Qi and S. C. Zhang, *Rev. Mod. Phys.,* 2011, **83**, 1057.

2 B. A. Bernevig and S. C. Zhang, *Phys. Rev. Lett.*, 2006, **96**, 106802.

3 C. L. Kane and E. J. Mele, *Phys. Rev. Lett.*, 2005, **95**, 226801.

4 M. Z. Hasan and C. L. Kane, *Rev. Mod. Phys.*, 2010, **82**, 3045.

5 B. A. Bernevig, T. L. Hughes, and S. C. Zhang, *Science,* 2006, **314**, 1757.

6 H. Weng, X. Dai, and Z. Fang, *Phys. Rev. X*, 2014, **4**, 011002.

7 C. L. Kane and E. J. Mele, *Phys. Rev. Lett.*, 2005, **95**, 146802.

8 C. Wu, B. A. Bernevig, and S. C. Zhang, *Phys. Rev. Lett.*, 2006, **96**, 106401.

9 C. Xu and J. Moore, *Phys. Rev. B*, 2006, **73**, 045322.

10 M. Konig, S. Wiedmann, C. Brune, A. Roth, H. Buhmann, L. W. Molenkamp, X. L. Qi, and S. C. Zhang, *Science*, 2007, **318**, 766.

11 I. Knez, R. R. Du, and G. Sullivan, *Phys. Rev. Lett.*, 2011, **107**, 136603.

12 Z. Liu, Z. F. Wang, J. W. Mei, Y. S. Wu, and F. Liu, *Phys. Rev. Lett.*, 2013, **110**, 106804.

13 L. Y. Li, X. M. Zhang, X. Chen, and M. W. Zhao, *Nano Lett.*, 2015, **15**, 1296.

14 Z. F. Wang, N. Su, and F. Liu, *Nano Lett.*, 2013, **13**, 2842.

15 Z. F. Wang, Z. Liu, and F. Liu, *Phys. Rev. Lett.*, 2013, **110**, 196801.

16 Z. Liu, C. X. Liu, Y. S. Wu, W. H. Duan, F. Liu, and J. Wu, *Phys. Rev. Lett.*, 2011, **107**, 136805.

17 C. C. Liu, S. Guan, Z. G. Song, S. A. Yang, J. B. Yang, and Y. Yao, *Phys. Rev. B*, 2014, **90**, 085431.

18 J. J. Zhou, W. Feng, C. C. Liu, S. Guan, and Y. Yao, *Nano Lett*., 2014, **14**, 4767.

19 F. C. Chuang, L. Z. Yao, Z. Q. Huang, Y. T. Liu, C. H. Hsu, T. Das, H. Lin, and A. Bansil, *Nano Lett.*, 2014, **14**, 2505.

20 M. Zhou, W. Ming, Z. Liu, Z. Wang, P. Li, and F. Liu, *Proc. Natl. Acad. Sci. USA*, 2014, **111**, 14378.

21 M. W. Zhao, X. Chen, L. Y. Li, and X. M. Zhang, *Sci. Rep.*, 2015, **5**, 8441.

22 A. H. Castro Neto, N. M. R. Peres, K. S. Novoselov, and A. K. Geim, *Rev. Mod. Phys.*, 2009, **81**, 109.

23 J. C. Charlier and S. Roche, *Rev. Mod. Phys.*, 2007, **79**, 677.

24 R. Rurali, *Rev. Mod. Phys.*, 2010, **82**, 427.

25 Y. Xu, B. Yan, H. J. Zhang, J. Wang, G. Xu, P. Tang, W. Duan, and S. C. Zhang, *Phys. Rev. Lett.*, 2013, **111**, 136804.

26 C. Si, J. Liu, Y. Xu, J. Wu, B. L. Gu, and W. Duan, *Phys. Rev. B*, 2014, **89**, 115429.

27 P. Tang, P. Chen, W. Cao, H. Huang, S. Cahangirov, L. Xian, Y. Xu, S. C. Zhang, and W. Duan, A. Rubio, *Phys. Rev. B*, 2014, **90**, 121408.

28 V. O. Özçelik and S. Ciraci, *J. Phys. Chem. C*, 2013, **117**, 26305.

29 S. Cahangirov, V. O. Özçelik, L. Xian, J. Avila, S. Cho, M. C. Asensio, S. Ciraci, and A. Rubio, *Phys. Rev. B*, 2014, **90**, 035448.

30 D. Kaltsas and L. Tsetseris, *Phys. Chem. Chem. Phys.*, 2013, **15**, 9710.

31 G. Kresse and J. Furthmüller, *Phys. Rev. B*, 1996, **54**, 11169.

32 G. Kresse and J. Hafner, *Phys. Rev. B*, 1993, **48**, 13115.

33 G. Kresse and D. Joubert, *Phys. Rev. B*, 1999, **59**,1758.

34 J. P. Perdew, K. Burke, and M. Ernzerhof, *Phys. Rev. Lett.*, 1996, **77**, 3865.

35 K. Parlinski, Z. Q. Li, and Y. Kawazoe, *Phys. Rev. Lett.*, 1997, **78**, 4063.

36 A. D. Becke and K. E. Edgecombe, *J. Chem. Phys.*, 1990, **92**, 5397.

37 A. Savin, O. Jepsen, J. Flad, O. K. Andersen, H. Preuss, and H. G. von Schnering, *Angew. Chem. Int. Ed.*, 1992, **31**, 187.

38 Z. Sun, C. L. Pint, D. C. Marcano, C. Zhang, J. Yao, G. Ruan, Z. Yan, Y. Zhu, and R. H. Hauge, *Nat. Commun.*, 2011, **2**, 559.

39 M. Pumera and C. H. Wong, *Chem. Soc. Rev.*, 2013, **42**, 5987.

40 S. Cahangirov, M. Topsakal, E. Aktürk, H. Sahin, and S. Ciraci, *Phys. Rev. Lett.*, 2009, **102**, 236804.

41 C. C. Liu, H. Jiang, and Y. Yao, *Phys. Rev. B*, 2011, **84**, 195430.

42 C. C. Liu, W. Feng, and Y. Yao, *Phys. Rev. Lett*., 2011, **107**, 076802.

43 V. O. Özcelik, E. Durgun, and S. Ciraci, *J. Phys. Chem. Lett.,* 2014, **5**, 2694

44 P. De Padova, P. Vogt, A. Resta, J. Avila, I. Razado-Colambo, C. Quaresima, C. Ottaviani, B. Olivieri, T. Bruhn, T. Hirahara, T. Shirai, S. Hasegawa, M. C. Asensio, and G. Le Lay, *Appl. Phys. Lett.,* 2013, **102**, 163106.

45 P. Vogt, P. Capiod, M. Berthe, A. Resta, P. De Padova, T. Bruhn, G. Le Lay, and B. Grandidier, *Appl. Phys. Lett.,* 2014, **104**, 021602.

46 B. Feng, Z. Ding, S. Meng, Y. Yao, X. He, P. Cheng, L. Chen, and K. Wu, *Nano Lett.*, 2012, **12**, 3507.

47 V. O. Özcelik, D. Kecik, E. Durgun, and S. Ciraci, *J. Phys. Chem. C,* 2015*,* **119***,* 845

48 M. Derivaz, D. Dentel, R. Stephan, M. C. Hanf, A. Mehdaoui, P. Sonnet, and C. Pirri, *Nano Lett.*, 2015, **15**, 2510.

49 J. Qiu, H. Fu, Y. Xu, A. I. Oreshkin, T. Shao, H. Li, S. Meng, L. Chen, and K. Wu, *Phys. Rev. Lett*., 2015, **114**, 126101.

50 L. Fu and C. L. Kane, *Phys. Rev. B*, 2007, **76**, 045302.

51 H. Zhang and S. C. Zhang, *Phys. Stat. Sol.*, 2013, **7**, 72.

52 Z. F. Wang, L. Chen, and F. Liu, *Nano Lett.*, 2014, **14**, 2879.

53 L.Y. Li and M. W. Zhao, *J. Phys. Chem. C*, 2014, **118**, 19129.

54 L.Y. Li and M. W. Zhao, *Phys. Chem. Chem. Phys.*, 2013, **15**, 16853.

55 L. Y. Li, X. P. Wang, X. Y. Zhao, and M. W. Zhao, *Phys. Lett. A*, 2013, **377**, 2628.

**Figure caption**

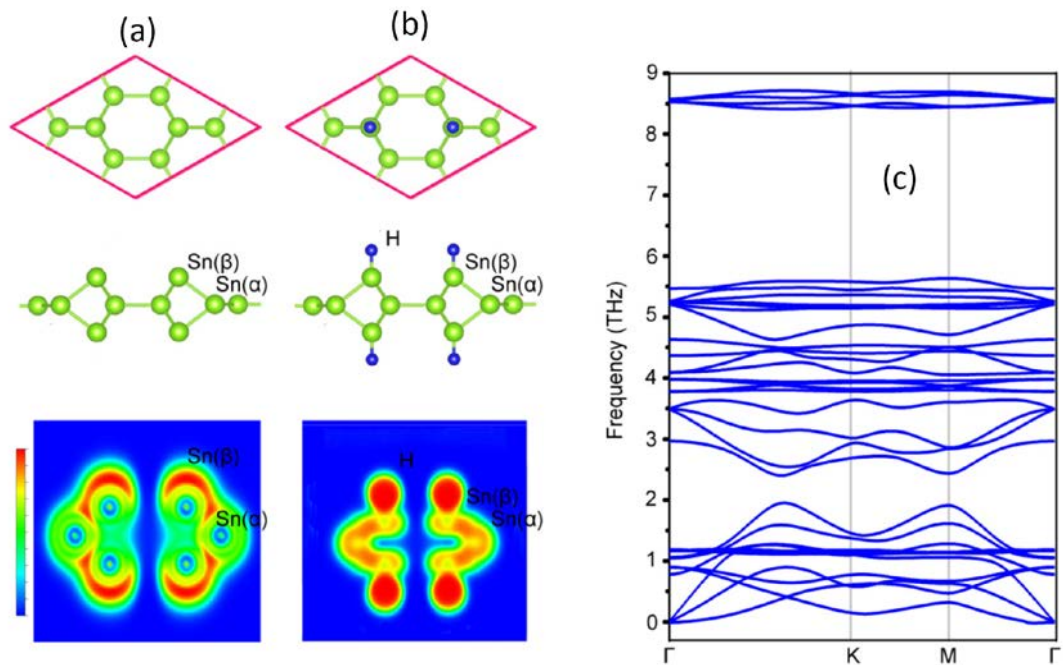

**Figure 1.** Top and side views (up panel) and the ELF profiles (down panel) of (a) DB stanene and (b) DB stanane. (c) Phonon spectra of DB stanane along the high-symmetric points in the BZ.

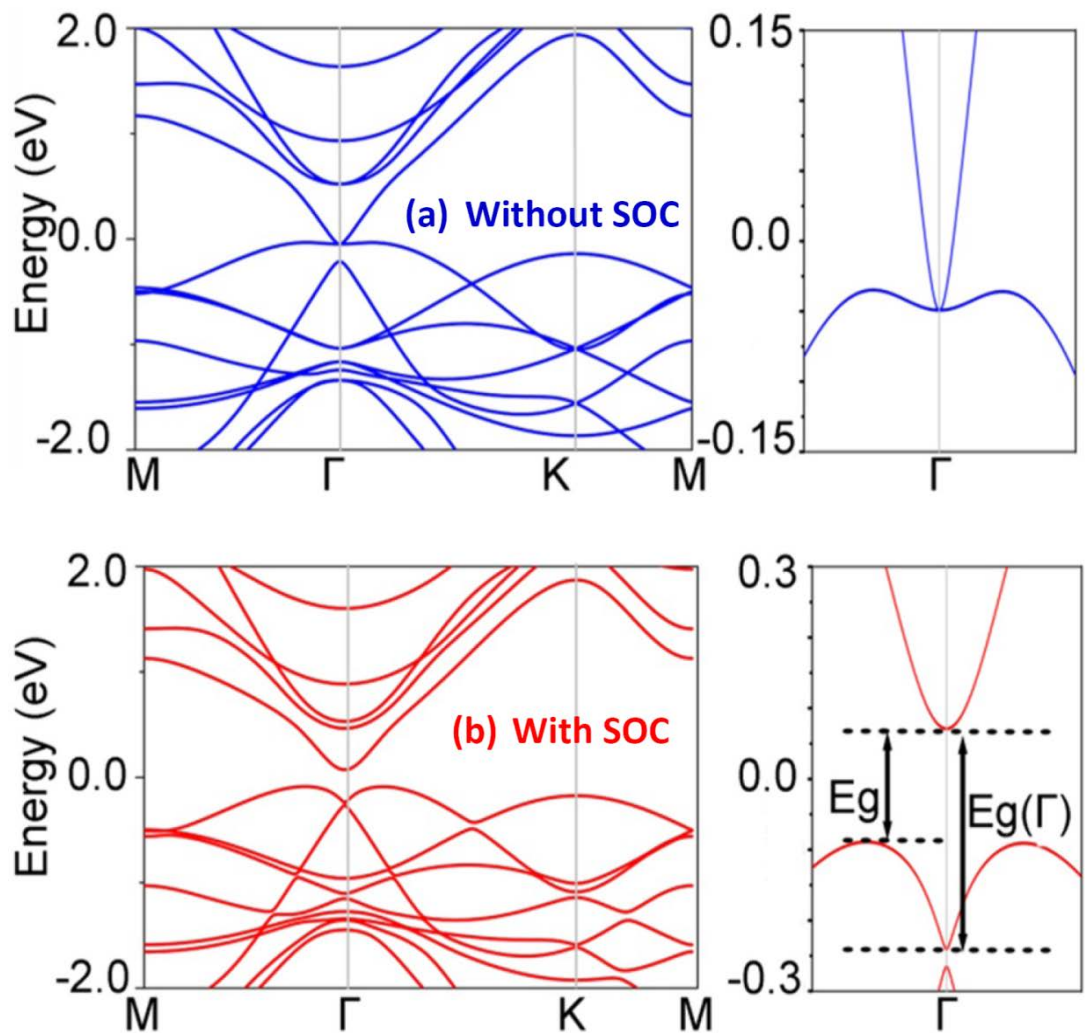

**Figure 2.** Band structures of DB stanane (a) without and (b) with SOC. The enlarged views of band lines near the Fermi level in the vicinity of the Γ point are shown in right panel of this figure. The energy at the Fermi level was set to zero.

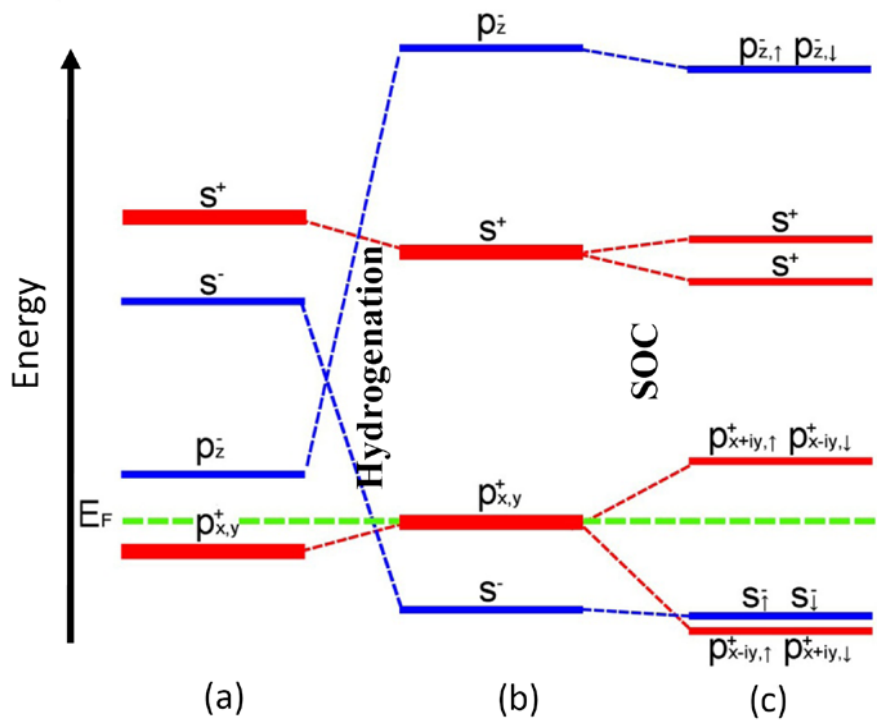

**Figure 3.** Schematic diagram of the evolution of the electronic states near the Fermi level of (a) DB stanene and (b), (c) DB stanane. The position of the Fermi level is denoted by the green line. The effects of hydrogenation and spin-orbit coupling (SOC) on the alignment of electronic states are indicated.

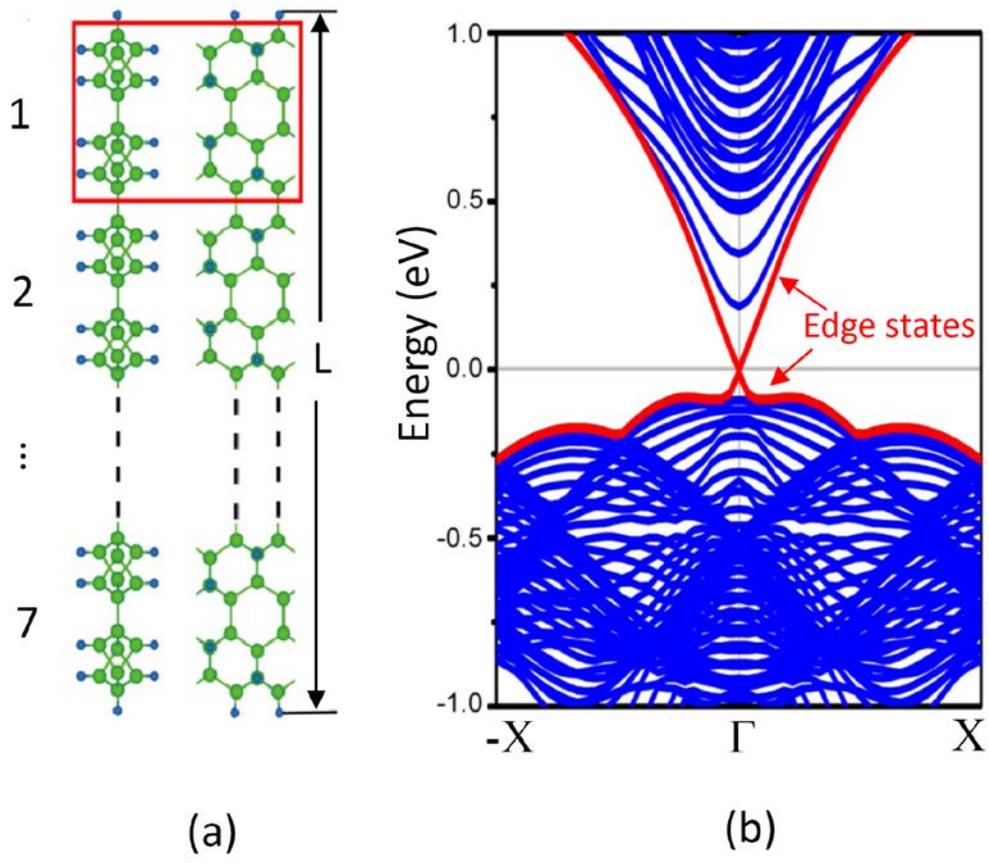

**Figure 4.** (a) Schematic representations (top and side views) of armchair-edged DB stanane with the width of L=10.9 nm. (b) Electronic band structure of the DB stanane nanoribbon. The helical edge states are indicated by the red lines. The energy at the Fermi level was set to zero.

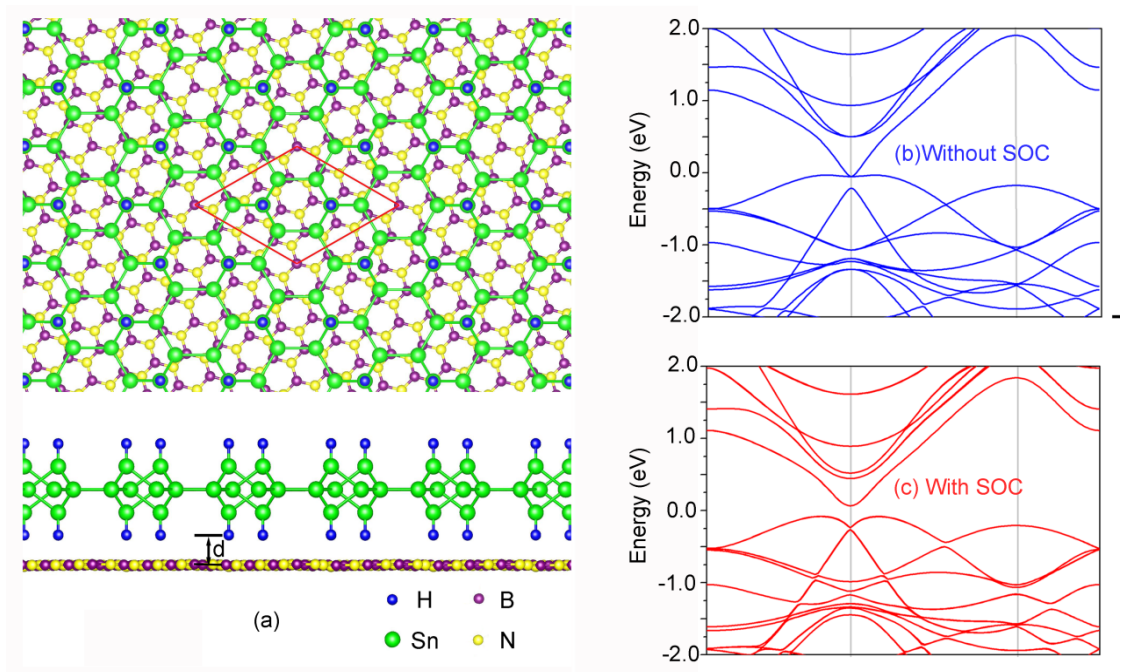

**Figure 5.** (a) Schematic representations (top and side views) DB stanane grown on hexagonal boron nitride (h-BN) substrate. The red rhombus indicates the unit cell of the system. Electronic band structures (b) without and (c) with SOC of the DB stanane on h-BN substrate. The energy at the Fermi level was set to zero.

Supplementary Information for "Dumbbell Stanane: A large-gap quantum spin Hall insulator"

Xin Chen, Linyang Li, Mingwen Zhao*

*School of Physics and State Key Laboratory of Crystal Materials, Shandong University, Jinan, Shandong, 250100, China*

E-mail: zmw@sdu.edu.cn

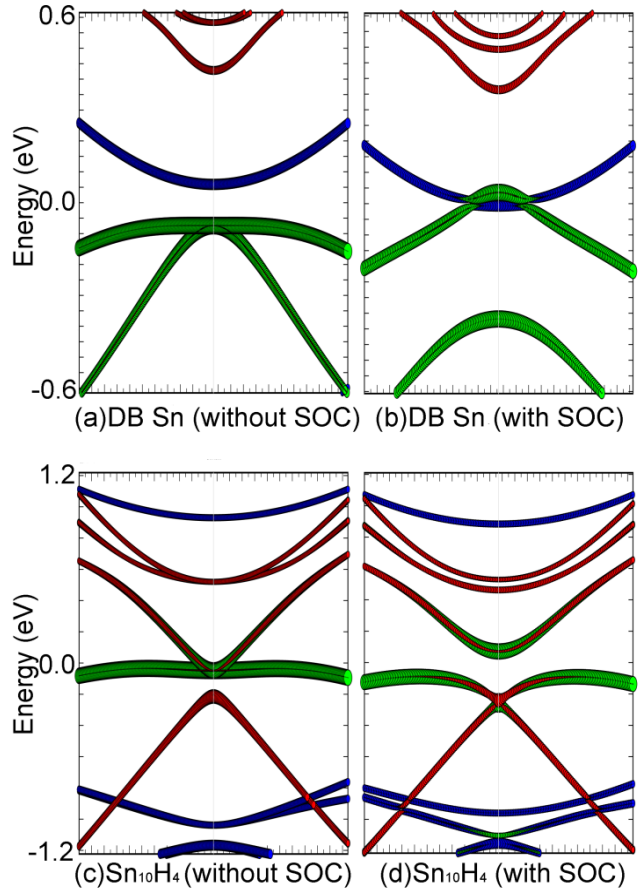

**Figure 1.** Orbital-resolved band structures of DB stanene and DB stanane around the Γ point from DFT calculations. The red dots represent the contributions from the s atomic orbital of Sn(α) atoms, the green dots represent contributions from $p_x$ and $p_y$ atomic orbitals of Sn(α) atoms, and the blue dots represent contributions from $p_z$ atomic orbitals of Sn(α) atoms. The $p_z$-$p_{xy}$ band inversion takes place in the pristine DB stanene owning to SOC while it is s-$p_{xy}$ band inversion in DB Stanane.